\documentclass{elsart}

\usepackage{amsmath,amsfonts,amssymb,bm}
\usepackage{graphicx,color,overpic,psfrag}
\usepackage{url}
\usepackage{cite}

\begin{document}

\begin{frontmatter}

\title{Security problems with a chaos-based deniable authentication scheme}

\author{Gonzalo Alvarez\corauthref{corr}}
\address{Instituto de F\'{\i}sica
Aplicada, Consejo Superior de Investigaciones Cient\'{\i}ficas,
Serrano 144---28006 Madrid, Spain}

\corauth[corr]{Email address: \texttt{gonzalo@iec.csic.es}.}

\begin{abstract}
Recently, a new scheme was proposed for deniable authentication.
Its main originality lied on applying a chaos-based
encryption-hash parallel algorithm and the semi-group property of
the Chebyshev chaotic map. Although original and practicable, its
insecurity and inefficiency are shown in this paper, thus
rendering it inadequate for adoption in e-commerce.
\end{abstract}

\end{frontmatter}

\section{Introduction}

In recent years, chaos-based cryptography is drawing a great deal
of attention from researchers from a variety of disciplines
\cite{Alvarez:Survey:ICCST99,
Kocarev:Chaos-Cryptography:IEEECASM2001,
Dachselt&Schwarz:Chaos&Cryptography:IEEETCASI2001,
ShujunLi:Dissertation2003, Yang:Survey:IJCC2004}. One of the most
interesting encryption algorithms based on chaos proposed to date
exploited the ergodic property of chaotic orbits
\cite{Baptista:CryptographyWithChaos:PLA98}. In the following
years, many other works enhanced or analyzed its speed and
security
\cite{Jakimoski:Analysis:PLA01,Wong:DynamicLookUpTable:PLA02,
Palacios:Cycling:PLA02, Wong:Hashing:PLA03,
LiShujun:ChaoticCipher:PLA2003,Alvarez:Ergodic:PLA03,
LiShujun:ChaoticCipher:PLA2004, Huang:ErgodicBinary:CSF05}. More
recently, a new scheme for deniable authentication making use of a
chaos-based encryption-hash parallel algorithm and the semi-group
property of the Chebyshev chaotic map was proposed
\cite{Xiao:DeniableAuthentication:CSF05}. In this paper it is
shown that the authors' claim to be ``secure and efficient'' may
be contradicted.

\section{The scheme}

According to \cite{Katz:PPK:EuroCrypt03}, the two main
characteristics of deniable authentication are: i) a sender
$\mathcal{S}$ (also called \emph{prover} in the literature) is
able to authenticate a message $m$ to a receiver $\mathcal{R}$
(also called \emph{verifier}); and ii) the receiver $\mathcal{R}$
is unable to convince a third party that a message $m$ was
authenticated by $\mathcal{S}$. An attacker $\mathcal{M}$ (acting
as man-in-the-middle between $\mathcal{S}$ and $\mathcal{R}$)
should not be able to authenticate a message $m$ to $\mathcal{R}$
which $\mathcal{S}$ does not authenticate for $\mathcal{M}$.

Many different constructions of deniable authentication protocols
have been published based on traditional cryptography (see for
example \cite{Katz:PPK:EuroCrypt03} and references therein).
Usually, these protocols require at a minimum a hashing algorithm
and a public key cryptography algorithm. The scheme proposed in
\cite{Xiao:DeniableAuthentication:CSF05} uses the chaos-based
encryption-hash parallel algorithm defined in
\cite{Wong:DynamicLookUpTable:PLA02,Wong:Hashing:PLA03} and the
Chebyshev chaotic map to realize key agreement, as proposed in
\cite{Kocarev:Chebyshev:ISCAS03}.

\subsection{Encryption-hash}

\begin{figure}[h]
\center \includegraphics{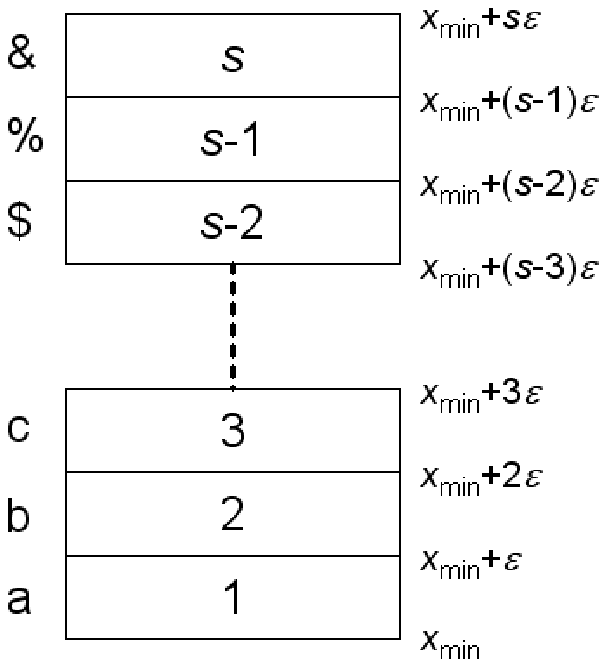}
\caption{\label{fig:ergodic}Schematic representation of how an
attractor is divided into $s$ subintervals, each one with size
$\epsilon = (y_{\max}-y_{\min})/s$. An alphabet unit is associated
for each subinterval.}
\end{figure}

The encryption-hash algorithm uses the logistic map
\[y_{n+1}=by_n(1-y_n),\]
where $y_n\in [0,1]$ and the parameter is $3.99<b<4.0$, so that it
behaves chaotically. Following
\cite{Baptista:CryptographyWithChaos:PLA98}, the interval
$[y_{\min},y_{\max}]$, where $0<y_{\min}<y_{\max}<1$, is divided
up into $s=256$ subintervals, in one-to-one correspondence to as
many ASCII characters (see Fig.~\ref{fig:ergodic}). The secret key
is given by the initial point $y_0$ and the parameter value $b$.
To encrypt an 8-bit block, i.e., an ASCII character, the orbit is
iterated starting from $y_0$ as many times as necessary until it
lands on the subinterval corresponding to the given ASCII symbol.
The number of iterations is recorded as the corresponding block
ciphertext. This procedure is repeated until the plaintext is
exhausted.

In \cite{Wong:DynamicLookUpTable:PLA02}, a dynamic table is used
for looking up the ciphertext and plaintext, which is no longer
fixed during the whole encryption and decryption processes as in
\cite{Baptista:CryptographyWithChaos:PLA98}. Instead, it depends
on the plaintext, being continuously updated during the encryption
and decryption processes. When the $i$th message block is
encrypted, the look-up table is updated dynamically by exchanging
the $i$th entry $l_i$ with another entry $l_j$ . The location of
the latter entry, i.e., the value of $j$, is determined by the
current value of $y$ using the following formula:
\[
\upsilon=\left\lfloor{\frac{{y-y_{\min}}}{{y_{\max}-y_{\min}}}}\times
s\right\rfloor,
\]
\[
j=i+\upsilon\mod s, \]
where $y_{\min}$ and $y_{\max}$ are the end
points of the chosen interval $[y_{\min},y_{\max})$ and $s$ is the
total number of entries in the table
\cite{Wong:DynamicLookUpTable:PLA02}.

In \cite{Wong:Hashing:PLA03}, the previously described chaotic
cryptographic scheme is generalized by allowing the swapping of
multiple pairs of entries in the look-up table during the
encryption of each input block, and by allowing multiple runs of
encryption on the whole message continuously. Starting from the
current entry $i$, $p$ pairs of entries ($p\geq 1$) are swapped
according to the following rule: $i\leftrightarrow(i+\upsilon\mod
s)$, $(i+\upsilon+1\mod s)\leftrightarrow(i+2\upsilon+1\mod s)$,
$(i+2\upsilon+2\mod s)\leftrightarrow(i+3\upsilon+2\mod s)$,
\dots, $(i+(p-1)\upsilon+p-1\mod s)\leftrightarrow
(i+p\upsilon+p-1\mod s)$. Once the message has been encrypted, the
whole process is repeated again $r$ times, $r\geq 1$. The final
look-up table is the hash of the message
\cite{Wong:Hashing:PLA03}.

\subsection{Session key agreement}

The key agreement protocol is based on Chebyshev polynomials and
their properties. The Chebyshev polynomial of degree $n$ is
defined as
\[T_n(x)=\cos(n\cdot\arccos(x)),\;x\in[-1,1].\]
The polynomial $T_n(x)$ is recursively defined as
\[T_{n+1}(x)=2xT_n(x)-T_{n-1}(x), \;\textrm{for any}\; n>0,\]
where $T_0(x)=1$ and $T_1(x)=x$. Chebyshev polynomials verify the
semi-group property: $T_p(T_q(x))=T_{pq}(x)$; and also commute
under composition: $T_p(T_q(x))=T_q(T_p(x))$. These two properties
make them eligible for public key cryptography and authentication
\cite{Kocarev:Chebyshev:ISCAS03}.

The key agreement process described in
\cite{Xiao:DeniableAuthentication:CSF05} is as follows:

\begin{enumerate}
    \item $\mathcal{S}$ and $\mathcal{R}$ choose a publicly known random number $x\in[-1,1]$.
    \item $\mathcal{S}$ chooses a random large integer $p$, computes
    $P=T_p(x)$ and sends $P$ to $\mathcal{R}$.
    \item $\mathcal{R}$ chooses a random large integer $q$, computes
    $Q=T_q(x)$ and sends $Q$ to $\mathcal{S}$.
    \item $\mathcal{S}$ computes the secret key as $k=T_p(Q)=T_p(T_q(x))$.
    \item $\mathcal{R}$ computes the secret key as $k'=T_q(P)=T_q(T_p(x))$.
\end{enumerate}

Due to the semi-group property, $k=k'=T_{pq}(x)$. All the
communication steps are susceptible to interception and
manipulation by an attacker: $x$, $P=T_p(x)$, and $Q=T_q(x)$ might
be known or altered by the attacker acting as a man-in-the-middle
$\mathcal{M}$. The security of this algorithm relies on the
assumption that given only the pair $(x,T_n(x))$ it is very
difficult to compute the order of the polynomial $n$.

\subsection{Deniable authentication protocol}

Once $\mathcal{S}$ and $\mathcal{R}$ have agreed on a common
session key $k$ as described before, $\mathcal{S}$ computes
$E_k(m)$ and obtains the hash value $H(m)$ simultaneously.
$\mathcal{S}$ sends $E_k(m)$ and $H(m)$ to $\mathcal{R}$. Now
$\mathcal{R}$ can decrypt $D_k(E_k(m))=m$ using the same session
key $k$, obtaining simultaneously the hash value $H'(m)$. If both
hashes $H$ and $H'$ are identical, $\mathcal{R}$ is assured that
the message $m$ was sent by $\mathcal{S}$. For a more thorough
description of the scheme, the reader is referred to the original
work \cite{Xiao:DeniableAuthentication:CSF05}.

\section{Analysis of the scheme}

In this section, the insecurity and inefficiency of the scheme
proposed in \cite{Xiao:DeniableAuthentication:CSF05} are analyzed.

\subsection{Security analysis of the scheme}

The security of the encryption-hash algorithm
\cite{Wong:DynamicLookUpTable:PLA02,Wong:Hashing:PLA03} was
already studied in \cite{Alvarez:BreakLookUpTable:PLA04}, where it
was showed that:

\begin{itemize}
    \item The algorithm is vulnerable to
chosen-ciphertext, chosen-plaintext and known-plaintext attacks.
As a consequence, implementations of this algorithm can never
reuse the same key because if so, they are easily broken.
    \item The look-up table, and thus the hash, does not depend on the key, but only on the plaintext, thus facilitating cryptanalysis.
    \item Breaking the hash algorithm is possible when $p=1$ and $r=1$, even without the knowledge
of the key $k$ ($y_0$ and $b$). In fact, it is easy to find two
different messages $m$ and $m'$ such that $H(m)=H(m')$.
\end{itemize}

These results imply that successive messages authenticated by
$\mathcal{S}$ should always use different session keys, thus
reproducing the key agreement protocol every time. This setting is
fundamental to avoid the attacks mentioned in the first bullet. In
order to avoid the type of attacks on the hashing scheme described
in the third bullet, it is all important that $r>1$ and $p>1$. Due
to the complexity of the attacks, the reader is referred to
\cite{Alvarez:BreakLookUpTable:PLA04} for a more detailed
explanation.

On the other hand, the security of the key agreement protocol was
studied in \cite{Bergamo:BreakChebyshev:arXiv}, where it was
showed that an attack permits to recover the corresponding
plaintext from a given ciphertext. The same attack can be applied
to produce forgeries if the cryptosystem used for signing
messages, as used in \cite{Xiao:DeniableAuthentication:CSF05}. The
weak spot of the protocol lies on the fact that there are several
Chebyshev polynomials passing through the same point. The attack
works as follows.

It is assumed that $\mathcal{M}$ knows $x$, $T_p(x)$ and $T_q(x)$,
which are publicly available in the communication channel between
$\mathcal{S}$ and $\mathcal{R}$. To get the secret key $k$:

\begin{enumerate}
    \item $\mathcal{M}$ computes a $p'$ such that $T_{p'}(x)=T_p(x)$.
    \item $\mathcal{M}$ recovers $k=T_{p'q}(x)=T_{p'}(T_q(x))$.
\end{enumerate}

Given $x$ and $T_p(x)$, it can be efficiently computed an integer
solution $p'$ to the equation $T_{p'}(x)=T_p(x)$:

\[p'=\frac{\pm\arccos(T_p(x))+2n\pi}{\arccos(x)}.\]

The reader is referred to \cite{Bergamo:BreakChebyshev:arXiv} for
the details on how to solve the previous equation, using a system
of two linear equations. This attack allows $\mathcal{M}$ to
actively forge a message from $\mathcal{S}$ to $\mathcal{R}$,
which makes the authentication property fail (Sec. 3.2.2 in
\cite{Xiao:DeniableAuthentication:CSF05}), or to passively decrypt
messages sent to $\mathcal{R}$ by $\mathcal{S}$, which makes the
security property fail (Sec. 3.2.3 in
\cite{Xiao:DeniableAuthentication:CSF05}).

\subsection{Efficiency analysis of the scheme}

Finally, in \cite{Xiao:DeniableAuthentication:CSF05} it is claimed
that the chaos-based encryption-hash parallel algorithm ``saves
certain computation time when compared with traditional hashing
and cryptographic methods''. This assertion might be interpreted
in the sense that their algorithm is faster than traditional
hashing and cryptographic methods, when in fact it is several
orders of magnitude slower. Table 1 of \cite{Wong:Hashing:PLA03}
gives some results to illustrate the performance of the proposed
chaotic cryptographic and hashing algorithm. The performance
depends on the values of $p$ and $r$. The best speed achieved is
between 7.7 and 11.5 KB/s in a 1.8 GHz processor. On the other
hand, traditional encryption algorithms, such as DES or AES,
achieve speeds of 21.3 and 61.0 MB/s respectively in a 2.1 GHz
processor \cite{Dai:SpeedComparison:URL}. With respect to
traditional hashing algorithms, MD5 and SHA-1, the two most widely
used, achieve speeds of 216.6 and 67.9 MB/s respectively in a 2.1
GHz processor \cite{Dai:SpeedComparison:URL}. Thus, the claim is
proved to be inadequate. As a consequence, this algorithm is also
very inefficient (between 1,000 and 10,000 times slower) when
compared to similar traditional algorithms.

\section{Conclusion}

The attacks proposed in \cite{Alvarez:BreakLookUpTable:PLA04} and
\cite{Bergamo:BreakChebyshev:arXiv} do not make the deniable
authentication protocol presented in
\cite{Xiao:DeniableAuthentication:CSF05} secure. An attacker can
forge messages in the name of the sender, thus violating the
authentication requirement, and can decrypt messages sent by the
sender, thus violating the security (confidentiality) requirement.
On the other hand, the use of an encryption-hash algorithm based
on discrete chaotic maps and on the ergodic property of chaotic
orbits greatly reduces the protocol speed, making it inefficient
as compared to other similar protocols. After these attacks, it is
concluded that the lack of security, along with the low operation
speed, may discourage the use of this scheme for secure
applications.

\ack{Thanks to Dr. Paolo D'Arco for his help and suggestions. This
work was supported by the Ministerio de Ciencia y Tecnolog\'{\i}a
of Spain, research grant TIC2001-0586 and SEG2004-02418.}

\clearpage\pagestyle{empty}

\section*{Figures}

\end{document}